\documentclass[conference]{IEEEtran}

\usepackage{amsmath,amssymb,amsfonts}
\usepackage{algorithm}
\usepackage{algorithmic}
\usepackage{graphicx}
\usepackage{textcomp}
\usepackage{ulem}
\usepackage{xcolor}
\usepackage{float}
\usepackage{booktabs}
\usepackage{hyperref}   

\usepackage{cite}

\def\BibTeX{{\rmB\kern-.05em{\sci\kern-.025emb}\kern-.08emT\kern-.1667em\lower.7ex\hbox{E}\kern-.125emX}}

\begin{document}

\title{Nano-Resolution Visual Identifiers Enable Secure Monitoring in Next-Generation Cyber-Physical Systems}


\author{\IEEEauthorblockN{Hao Wang\textsuperscript{1}, 
Xiwen Chen\textsuperscript{1}, 
Abolfazl Razi\textsuperscript{1},
Michael Kozicki\textsuperscript{2}, 
Rahul Amin\textsuperscript{3} and 
Mark Manfredo\textsuperscript{4}}
\IEEEauthorblockA{\textsuperscript{1}School of Computing, Clemson University }
\IEEEauthorblockA{\textsuperscript{2}School of Electrical, Computer and Energy Engineering, Arizona State University}
\IEEEauthorblockA{\textsuperscript{3}MIT Lincoln Lab} 
\IEEEauthorblockA{\textsuperscript{4}Morrison School of Agribusiness, Arizona State University}
}

\maketitle

\begin{abstract}

Today's supply chains heavily rely on cyber-physical systems such as intelligent transportation, online shopping, and E-commerce. It is advantageous to track goods in real-time 
by web-based registration and authentication of products after any substantial change or relocation. 
Despite recent advantages in technology-based tracking systems, most supply chains still rely on plainly printed tags such as barcodes and Quick Response (QR) codes for tracking purposes. Although affordable and efficient, these tags convey no security against counterfeit and cloning attacks, raising privacy concerns. It is a critical matter since a few security breaches in merchandise databases in recent years has caused crucial social and economic impacts such as identity loss, social panic, and loss of trust in the community.

This paper considers an 
end-to-end system using dendrites as nano-resolution visual identifiers to secure supply chains. Dendrites are formed by generating fractal metallic patterns on transparent substrates through an electrochemical process, which can be used as secure identifiers due to their natural randomness, high entropy, and unclonable features.
The proposed framework compromises the back-end program for identification and authentication, a web-based application for mobile devices, and a cloud database. We review architectural design, dendrite operational phases (personalization, registration, inspection), a lightweight identification method based on 2D graph-matching, and a deep 3D image authentication method based on Digital Holography (DH). A two-step search is proposed to make the system scalable by limiting the search space to samples with high similarity scores in a lower-dimensional space. We conclude by presenting our solution to make dendrites secure against adversarial attacks.

\end{abstract}


\begin{figure}[]
    \centering
    \includegraphics[width=1\columnwidth]{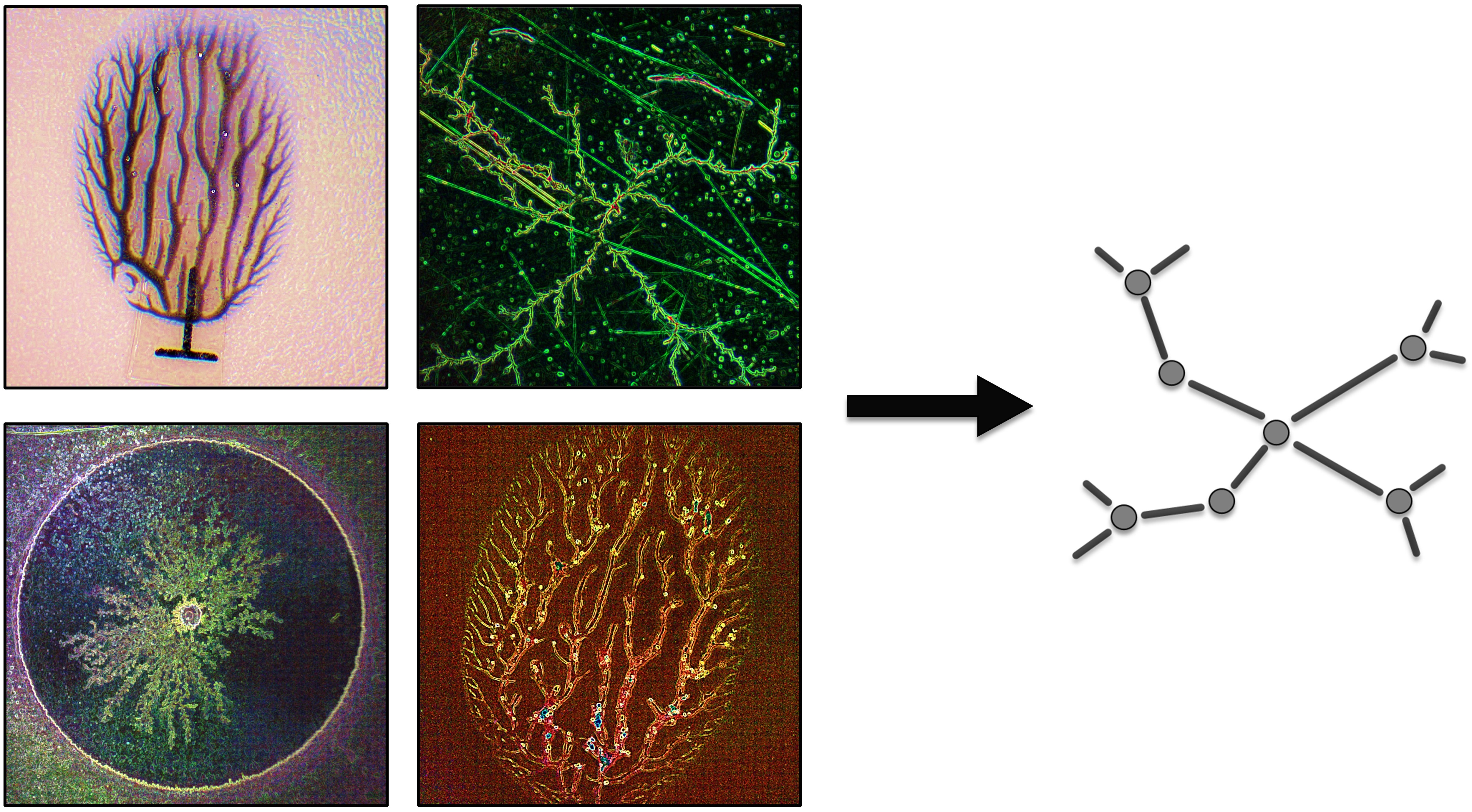}
    \caption{Using dendrite samples for optical authentication. \textbf{(left)} different classes of dendrite samples, \textbf{(right)} graph representation of a dendrite.}
    \label{fig:sample}
\end{figure}

\begin{figure*}[h]
    \centering
    \includegraphics[width=1\textwidth]{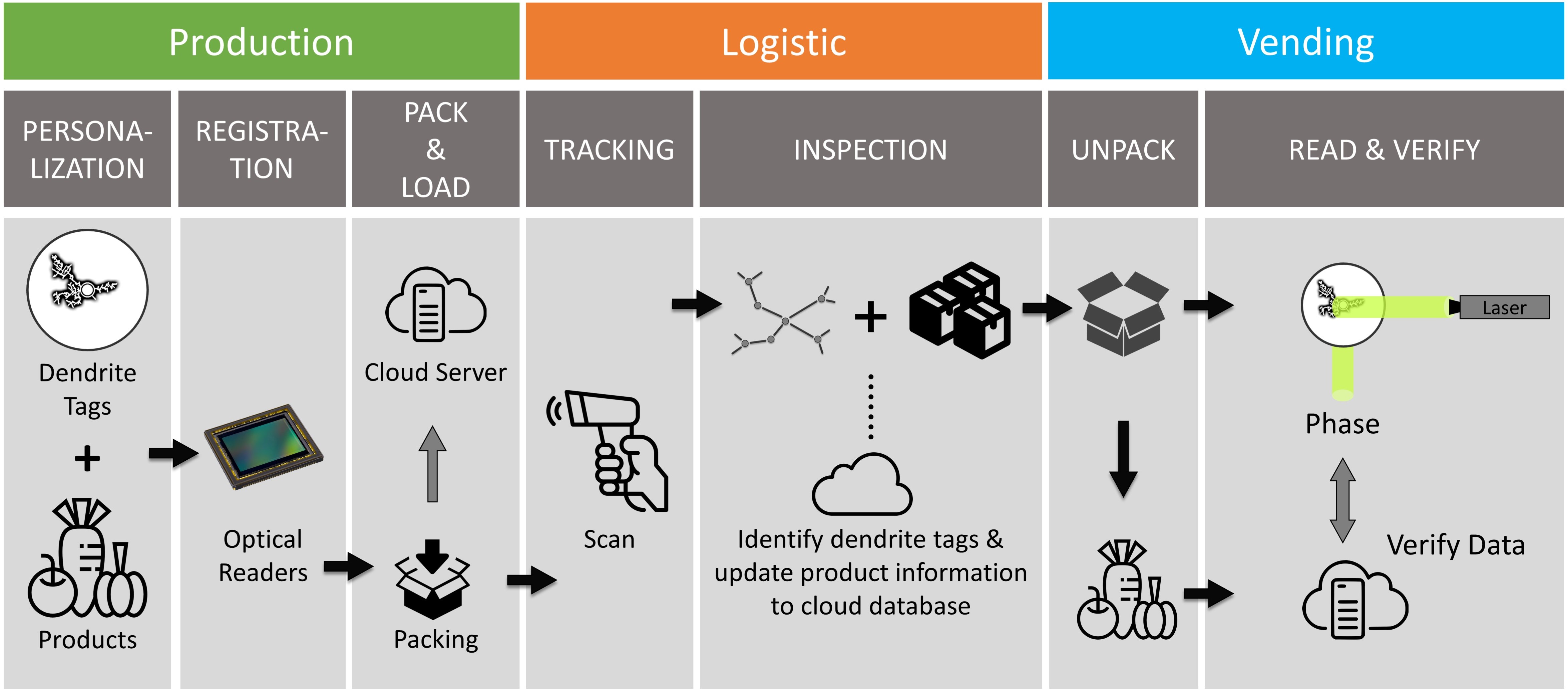}
    \caption{The overall workflow of the dendrite-based supply chain tracking and tracing. It shows different phases a denrite tag undergoes during its life cycle.}
    \label{fig:framework}
\end{figure*}

\section{Introduction}
Real-time tracking is becoming more and more popular these days. The essence of most tracking systems is associating location and time tags for the target's activity. In the modern era of smart cities with ubiquitous wireless access, real-time monitoring can be performed by web-based enterprises through tracing human's and smart IoT nodes' web activities. For passive objects, it can be realized by frequent registration of the object when its location changes (like in mail tracking systems). However, most tracking systems lack any sort of security making them vulnerable to counterfeit, manipulation, and physical attacks. For instance, one may take a copy of a FedEx shipment label and use it to send a hazardous or fake item to a target location. This paper considers using secure visual identifiers that are not clonable with existing technology.

Identification techniques are widely used in logistics applications, transportation, and COVID-19 pandemic monitoring \cite{brough2021consumer}, which is expected to provide rich, accurate, and secure information for tracing and tracking purposes. Some applications have relatively strict safety and privacy concerns \cite{burmester2007rfid}, and compromising security can lead to catastrophic consequences.   
For example, counterfeit, fake, adulterated, or falsified medicine and pharmaceutical products can result in severe and unpredictable risks to public health, especially when hospitals and emergency care providers are overwhelmed, such as Covid-19 pandemic \cite{ziavrou2022trends}. According to  \cite{ofori2022fighting}, the global black market industry 
accounts for 200 to 432 billion dollars annually. More seriously, the Drug Enforcement Administration (DEA) and Department of Justice recorded over 100,000 drug overdose deaths during 2021, and the primary driver is counterfeit Fentanyl \cite{counterfeit_pills_fact_sheet}. One reason for this fact is that a typical pharmaceutical supply chain is complex and fragile. It often consists of multiple agencies, such as the government, hospitals, drug manufacturers, drug distributors, and retailers. This means criminals can counterfeit the identification of drugs by misrepresenting information and/or falsifying the identity of unauthenticated products at each stage. This phenomenon can also happen in other supply chains, such as military, agriculture, and law enforcement, compromising public safety, food safety, and social stability. Therefore, an improved tagging and identification technique with privacy-preserving is necessary to secure such applications.

Although barcodes and Quick Response (QR) codes facilitate low-cost, efficient, and cellphone-readable tagging for identification and monitoring daily products, they apparently are easy targets for cloning attacks \cite{krombholz2014qr}. The biometrics-based identification approaches have better security; however, they are applicable only to humans and handheld items, not general objects. 
Recently, Radio Frequency Identification (RFID) tagging techniques have been used in some tracking applications (such as EZpass) due to their advantages, such as high successful identification rate, ease of implementation and trace, and remote access 
\cite{tan2022review}. This approach is relatively more secure than code-based approaches, but still suffers from different types of attacks such as cloning attack \cite{huang2020acd}, side-channel attack \cite{xu2018side}, etc. Moreover, developing a custom-built reader is relatively costly. Therefore, we consider a visual identifier \textit{dendrite} that is developed by our team and has the following benefits:  
\begin{itemize}
    \item The identifier contains rich 2D/3D information, and its pattern is infeasible to clone.
    \item It is cheap and can be printed on a variety of substrates, making it appropriate for massive production.
    \item It can be printed with bio-safe material, so appropriate for farm and eatable products 
    \item The granularity, resolution, and overall application-oriented security features are easily controllable in the production line
    \item Can be inspected by a cellphone camera with internet connectivity.
\end{itemize}

A dendrite sample (shown in Fig. \ref{fig:sample}) generally consists of a fractal metallic pattern which can be used to construct Physical Unclonable Functions (PUF) to improve authentication systems \cite{kozicki2021secure}. Due to their special appearance, dendrites can be authenticated using graph-theoretic solutions \cite{chi2020consistency}. The depth information under its surface is more complicated that can only be retrieved using specific technologies such as digital holography \cite{li2020deep, chen2022dh}, making it almost impossible to duplicate with existing technology. The choice of substrate is quite wide and includes mica, metallic substrates, transparent, and even regular papers. Dendrite production is highly efficient and easily scalable to massive production \cite{kozicki2022dendritic}. Thus, compared to other authentication methods for highly secure applications \cite{thiyaneswaran2020development, kala2021contactless}, dendrites arise as an ultra-low-cost, size flexible, privacy-friendly, highly secure, and unclonable technology. Such patterns can be utilized by countless applications ranging from personal identification, disease tracking to facility security, criminal detection, supply chain monitoring, and Internet of Things (IoT) integrity \cite{ekiz2021end,kumari2021novel}.

In this paper, we present a dendrite-based identification system design from end-user to back-end enterprise (Fig. 2). 

\section{Methodology}

\subsection{System Design}
In contrast to conventional secure systems, the proposed framework emphasizes simplicity by focusing only on the efficiency of the identification and authentication processes. 
Fig. \ref{fig:framework} shows the conceptual framework and operational steps of the proposed system.
In general, we use dendrite tags as media for identification and authentication. 

\begin{figure}[h]
    \centering
    \includegraphics[width=1\columnwidth]{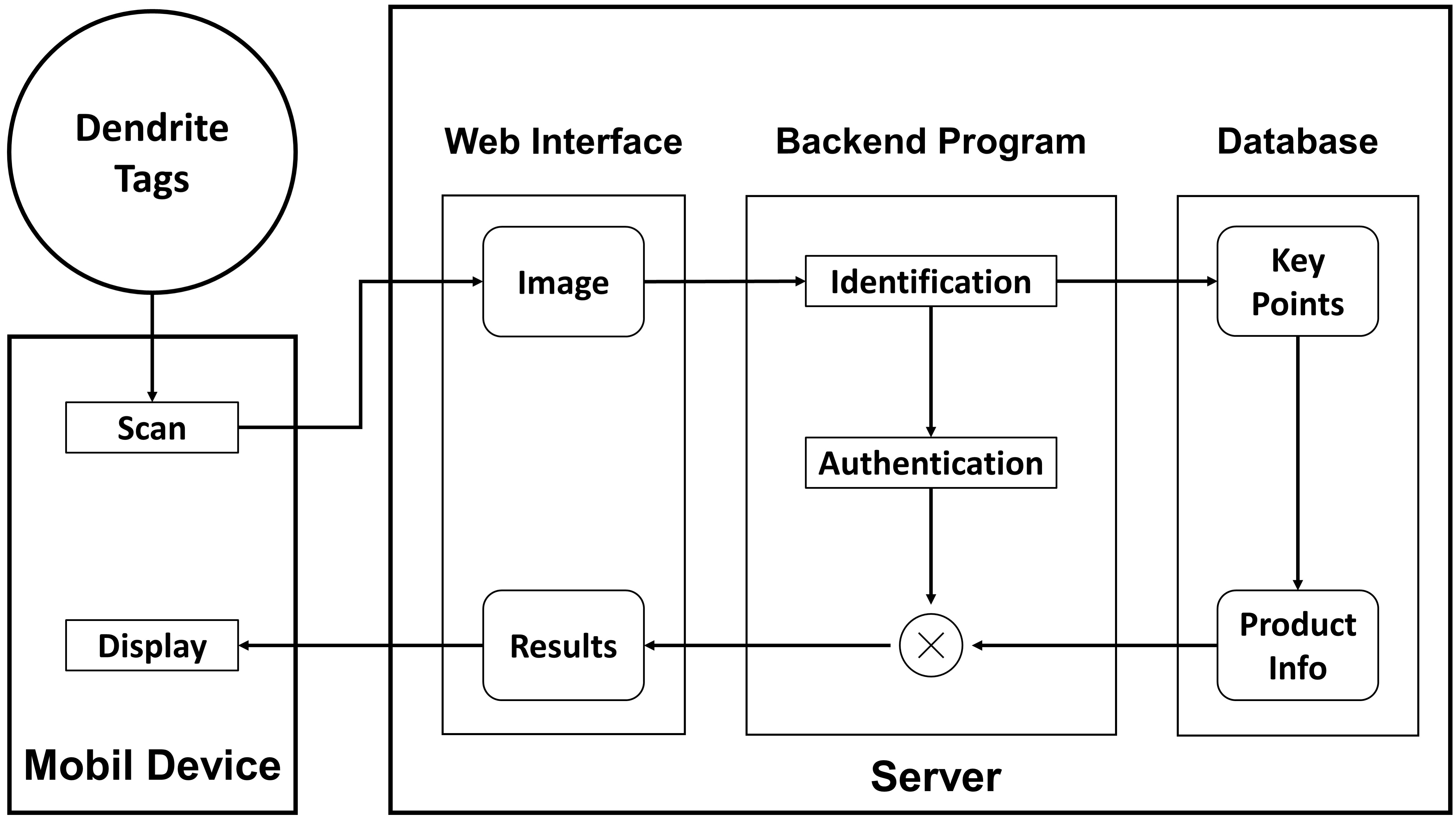}
    \caption{Cloud-based Authentication Center. The cloud server contains three main parts: a web interface, a back-end program, and a database.}
    \label{fig:cloud}
\end{figure}

In the production process, we assign each product a unique dendrite tag. 
The dendrite supplier uses high-resolution optical sensors (camera, scanner, etc.) to record the 2D morphological features of the tags, which are converted to a unique ID using a custom-built graph-theoretic algorithm \cite{wang2022fast}. At the entry point of the supply chain, the sample is attached to the product (by the product manufacturer). Simultaneously, the ID is associated with the product information in a cloud database (registration). Therefore, each record in the authentication center has two parts, dendrite features and product information.  

When the product is in transit from the supplier to the end-user, the tag can be scanned to authenticate the product, retrieve the logistic information and update the tracking history. If authentication is successful, the temporal data will be synchronized within the cloud server. The logistic data may contain tracking and tracing information, and other important information such as date, location history, product status, etc. Any unsuccessful authentication raises a red flag and triggers the security inspection module. All authentication attempts are recorded in the product activity log record (for further investigations).
Finally, when the product has been delivered to the end-user, the dendrite tag can be used for timely identification of the product and retrieving the product information. 
This prevents penetrating fake and replaced products during transit, as a key requirement of critical supply chains such as medicines and military logistics. We have implemented two classes of algorithms for authentication that includes 2D graph-matching \cite{valehi2017graph} as an affordable and low-complexity for general applications as well as a 3D nano-resolution authentication based on digital holography \cite{chen2022dh} for more secure applications.

\subsection{Cloud based Platform}
Different product vendors spent billions of dollars on preventing their products and database from hardware and software attacks \cite{lee2021cybersecurity}. Due to the high maintenance cost of local servers and the proven vulnerability of localized systems, cloud-based system designs are more desirable. 
In this work, a cloud-based platform has been proposed to store object information and track them through the supply chain. Compared to the fixed localized database, the cloud platform is low-cost and flexible that can be widely deployed to suppliers, delivery centers, and vendors. The proposed system consists of \textbf{three} main parts: a web-based application, a back-end program (consisting of an identification system and an authentication system), and a database. 
These modules work in harmony to prevent cyber-attacks. The web-based user interface is cross-platform compatible with Windows, Linux, iOS, Android, etc.

Fig. \ref{fig:cloud} presents a conceptual block diagram of the proposed cloud platform. Generally, a mobile device takes a picture of a dendrite sample and sends it to the web interface. The input image is processed by the back-end program, and the extracted key points information is passed to the database. The database returns the corresponding dendrite information in terms of key points that collectively determine a representative graph. 
Meanwhile, the back-end program verifies the input image using any of the 2D and 3D information. Note that we have tested our 3D DH reader, and our team is working on designing a portable version. After the image has been verified, the server will return the product information and the validation results to the mobile device. Thus, the product information has been retrieved by its dendrite tag.

\begin{figure}[h]
    \centering
    \includegraphics[width=1\columnwidth]{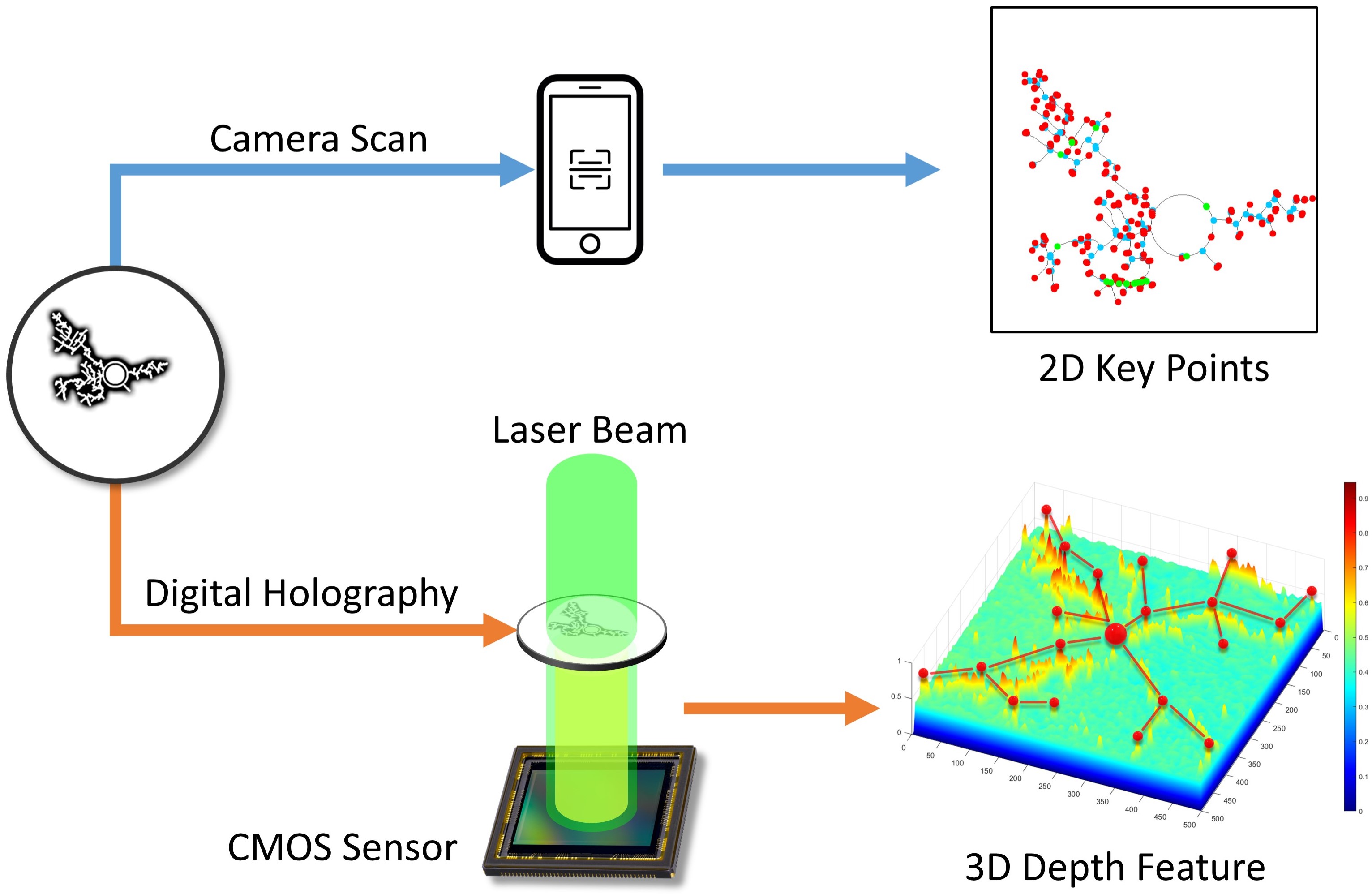}
    \caption{Two modes of dendrite authentication: (1) fast and affordable authentication using key points extraction and graph matching with regular cellphone cameras, (2) 3D features exploitation and 3D matching using laser-based digital holography for boosted security.}
    \label{fig:2d3d}
\end{figure}

\begin{figure*}[h]
    \centering
    \includegraphics[width=0.8\textwidth]{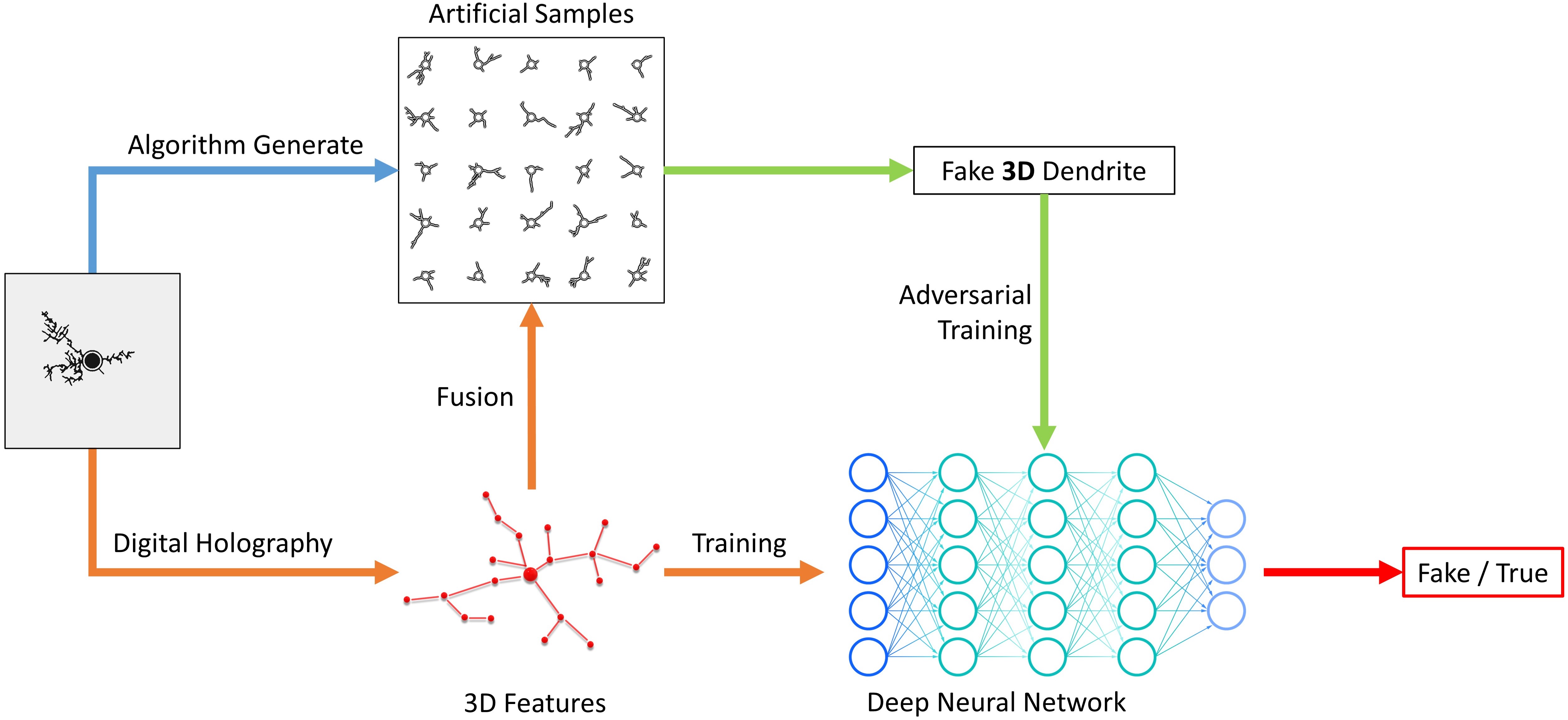}
    \caption{The architecture of a deep learning-based neural network for adversarial attacking defense.}
    \label{fig:attack}
\end{figure*}

\subsection{2D Graph Matching-based Identification}\label{sec:ident}

Compared to conventional image-based identification systems such as QR codes and barcodes, the dendrite patterns contain richer planar information, making it high-entropy. In contrast to QR codes, encoding (and interpreting) dendrite patterns is flexible, so it can be kept secret from the public and might be different from one application to another. For instance, one may extract key points within a certain angular perspective and up to a certain tree depth. The nano-resolution 3D facet makes a dendrite barely clonable.
For example, a simple copy of the QR code and barcode can be done by a regular printer. However, the copy process of the dendrite is much more complicated and even impossible due to the two-fold advantages. First, the resolution of the dendrite can be infinitely small due to its nano-resolution properties. 
Secondly, the generated pattern is high-entropy and fully random, making it impossible to reproduce the same pattern, even using the same equipment and under exactly the same conditions.

In \cite{wang2022fast}, we applied a fast learning-based key point detection algorithm to extract the tree-like graph of a dendrite image. The proposed method is fast and can be applied to similar tasks. As shown in Fig. \ref{fig:2d3d}, key points information can be extracted from the dendrite image scanned by mobile devices (for 3D features, special adapters for polarized imaging or DH is required). The extracted key points can be used for identification by matching the key points of the sample dendrite against a set of stored dendrite images.

Nevertheless, the image-to-image comparison is extremely inefficient. Comparing features to features is also time-consuming when applied to a 10,000-scale database. Thus, using the 2D feature-based dataset for fast identification can be desirable in practice.

Conventional password management systems use a dictionary data structure to store data and perform authentication by directly comparing the plain text, which is unsecured and easy to attack. 


In this work, we apply feature compression to achieve fast and safe identification. Specifically, we apply dimension reduction methods such as Principal/Independent Component Analysis (PCA/ICA) or data-driven methods to project high-dimension features into a low-dimension plane (e.g., 2D plane), and use the lower-dimension location to record surrogate feature information. The identification can be performed in two steps. First, we identify a set of close neighbors based on the pairwise distance between the lower-dimensional surrogates of the test and reference samples. Then, we execute the more complex graph-matching algorithm to identify the right match among the close neighbors. Dimension reduction is a one-way operator. Also, the projection depends on the collection of samples. Therefore, original patterns (or features) are not directly obtainable from the surrogate samples. Hence, this data is less sensitive and does not need to be encrypted. 
Overall, a low-complexity algorithm is executed for all reference samples in the database only once in the system's lifetime. During the test phase, the high-complexity deep matching of a test sample is performed with respect to a fixed number of closer reference samples. Therefore, the system is fast, safe, and easily scalable.

\subsection{Deep Authentication using 3D Image Shape Reconstruction}
Identification and authentication with reasonable confidence can be achieved by the proposed 2D graph matching. However, for enhanced security against cloning attacks, we rely on the 3D shape reconstruction of dendrites due to their abundant depth information. Note that 3D information can only be extracted through specific technologies such as tomography \cite{kalender2006x}, digital holography \cite{chen2022dh}, and polarization imaging \cite{abrahamsson2015multifocus}. In \cite{chen2022dh}, we proposed a DL-based method to retrieve depth information of digital holograms by recovering the phase of complex-valued laser wavefront. As shown in Fig. \ref{fig:2d3d}(b), the retrieved phase information contains a depth map that shows how the density/height of the metallic material is propagated through the dendritic pattern.

Although well-trained DL models can be used to attack advanced optical PUFs \cite{rivenson2018phase}, our proposed 3D reconstruction system in \cite{chen2022dh} still retains its advantages since the reconstruction of digital holograms requires either a training dataset (which we do not utilize in our untrained DL method) or it requires full knowledge of the exact system configuration parameters and the physics laws of laser back-propagation (which is not available). Thus, our proposed 3D reconstruction method is robust to DL attacks by data-driven models.

\subsection{Defence against Adversarial Attacks}
Despite the unclonable feature of dendrite samples, it is possible that a DL model can perform an adversarial attack by generating fake dendrite samples with synthetic 3D structures that can fool the authentication system \cite{yang20173d}. 
Therefore, an additional security level can be implemented to filter out fake samples even before undergoing the two-level authentication. 
To this end, we propose a neural network-based method for false sample detection, as shown in Fig. \ref{fig:attack}. Specifically, we first apply digital holography for dendrite samples to retrieve their depth information. Meanwhile, we generate artificial dendrite samples using mathematical algorithms. Then, we fuse the depth features with the artificial samples to generate fake dendrite samples with depth information. Finally, we train a discriminate network using both true samples and fake samples for false detection (classification). This method is not currently necessary because there exists no dendrite copy technology today, but it can be used to protect the system against cloning attacks by next-generation fabrication technology as well as emerging quantum-based adversarial attacks.

\begin{figure}[h]
    \centering
    \includegraphics[width=1\columnwidth]{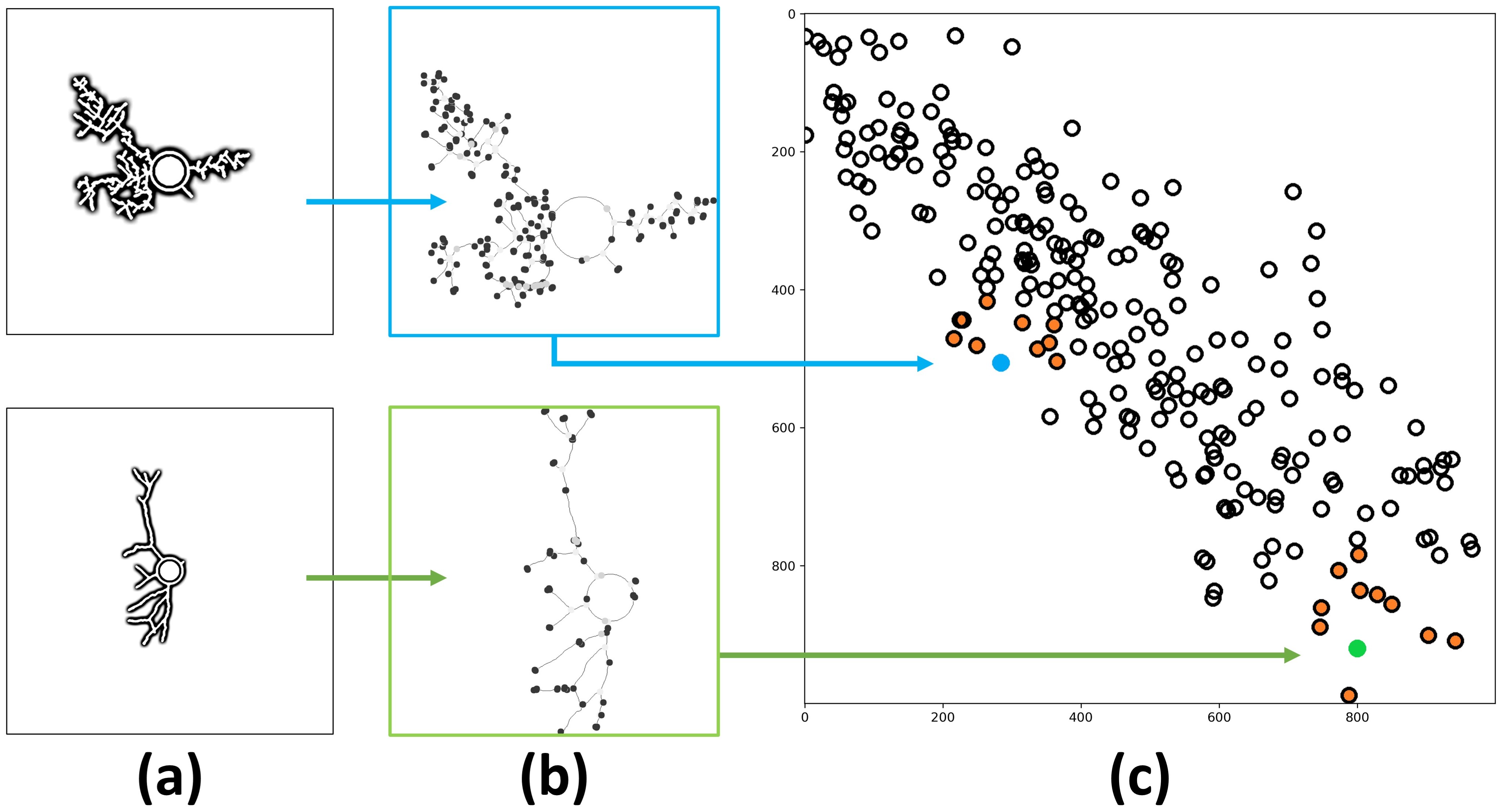}
    \caption{Identification process using PCA: (a) original dendrite image, (b) extracted key points, (c) the location of dendrite samples in 2D plane, test samples (blue and green dots) are compared only to its nearest neighbors (orange dots). }
    \label{fig:pca}
\end{figure}

\section{Results}

\subsection{Identification}

As discussed in Section \ref{sec:ident}, we proposed a 2-step sequential search algorithm based on limiting the search space to similar tags in lower dimensional space (instead of using the entire reference dataset), before running the relatively high-complexity deep authentication algorithm. The process is shown in Fig. \ref{fig:pca}.
The laboratory results show that the proposed identification system takes only $0.282$ second to match a new sample among 3,000 samples. This suggests that the proposed framework is highly efficient and scalable, qualifying it for secure tracking and tracing of large-scale supply chains and other applications.  

\begin{figure}[]
    \centering
    \includegraphics[width=1\columnwidth]{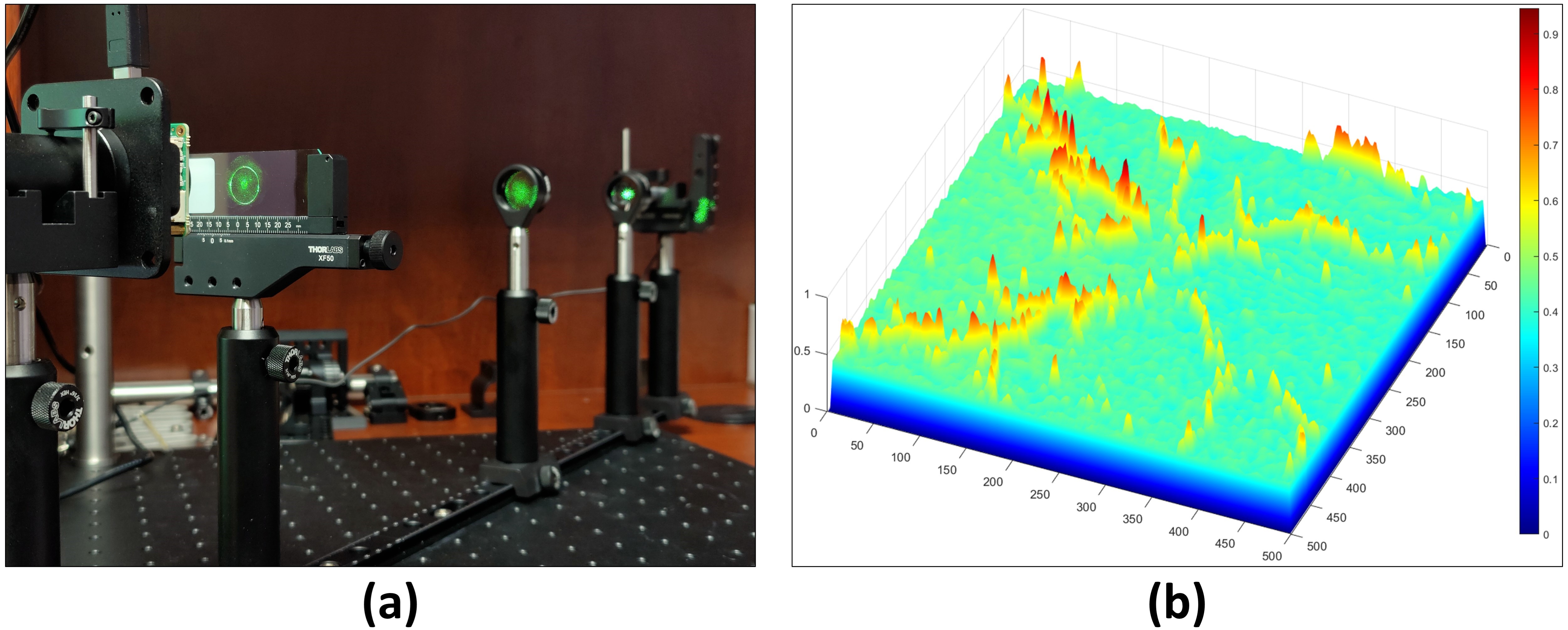}
    \caption{The proposed 3D reconstruction method based on digital in-line holography. (a) digital holography, a collimated laser beam pass through a sample and project its phase information into a CMOS sensor, (b) retrieved depth information from the captured hologram using a deep neural network.}
    \label{fig:holo}
\end{figure}


\subsection{Digital Holography for 3D Inspection}
Digital holography is a flexible technique used for reconstructing the 3D shape of microscopic objects \cite{amann20193d}. 
In \cite{chen2022dh}, we implemented a deep learning method based on an affordable DH platform (\$800) for 3D reconstruction. 
Specifically, our proposed method uses digital in-line holography to capture the hologram of samples (Fig. \ref{fig:holo}(a)), then we apply a Generative Adversarial Network (GAN)-based deep learning model to recover the phase information from the captured holograms (Fig. \ref{fig:holo}(b)).
It proves that digital holography is an appropriate technology to acquire the depth information of nano-scaled patterns. More importantly, the proposed method in \cite{chen2022dh} is training-free and suitable for transfer learning, which guarantees the fast deployment of 3D imaging in various applications.

\subsection{Software Architecture}
\begin{figure}[h]
    \centering
    \includegraphics[width=1\columnwidth]{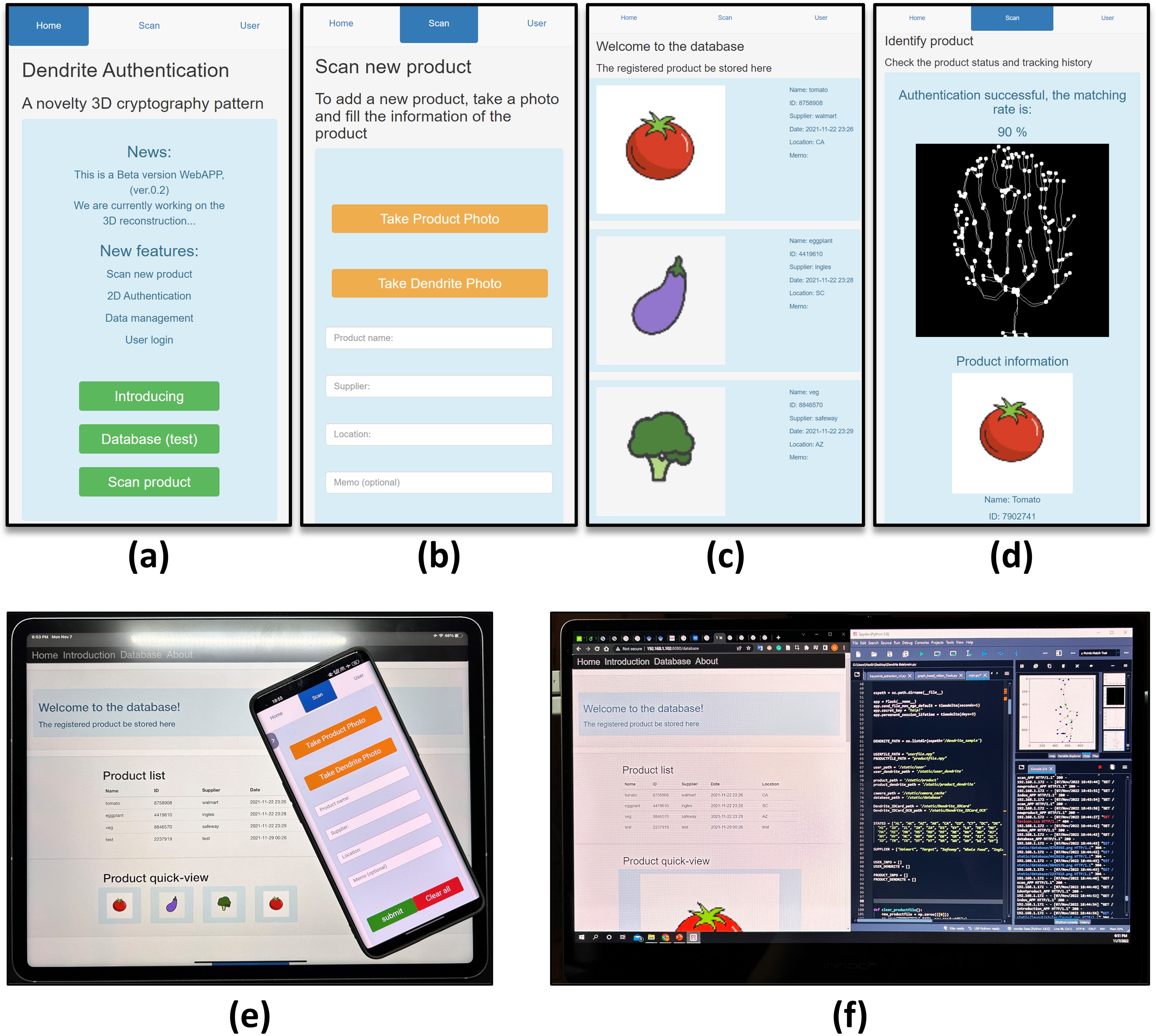}
    \caption{A view of the proposed web-based application when running on mobile devices. (a) The welcome page and main menu, (b) the new product registration page, (c) a database quick view, (d) the identification result after scanning a product, (e) an illustration of the cell phone App, (f) using a laptop to manage the back-end program and the cloud database.}
    \label{fig:app}
\end{figure}

The cloud server comprises a database, a back-end program, and a web-based application for mobile devices.
Fig. \ref{fig:app} shows the sample functions of our currently developed cloud-based platform when running on cell phones. The proposed platform demonstrates functions such as new product registration, product identification, and product tracking and management. 

As a new product is registered, a random dendrite tag is assigned to this product (personalization). Then, the product information such as supplier, date, location, and dendrite key points is stored in the database on the cloud server. When a product is in transit (from the producer to the end-user), it can be scanned by authorized personnel to update tracking history. 
When a product is scanned by vendors, the dendrite tag is used for identification by comparing its key points information with the database, and the product information and authentication result is returned to the vendor.

\section{Conclusion}
In this paper, we presented the system-level design of an end-to-end authentication platform that uses dendritic nano-scaled patterns for 3D authentication of objects. 
The system comprises a graph theoretic 2D image identification method for inspecting visual tags for identification-based tracking and tracing, a digital in-line holography-based deep learning model for deep authentication of dendritic tags by exploiting their 3D information, and a cloud-based server for image processing and data management. We proposed using surrogate sample matching in a lower dimensional space to find closed matches and accelerate the subsequent deep authentication. 

Due to the properties of massive productibility, low cost, bio-safety, device friendly, and easy integration with web-based services, the proposed system can be easily deployed to provide secure transit for supply chain products and other secure-sensitive goods such as medicine and military equipment. 

Our results show that the proposed framework is highly efficient, and scalable qualifying it for the next-generation cyber-physical systems 
to protect objects in zero-trust environments against the rising artificial intelligence-based and quantum computing-assisted security attacks.



\bibliography{ref}

\begin{thebibliography}{10}

\bibitem{brough2021consumer}
Brough AR, Martin KD.
\newblock Consumer privacy during (and after) the COVID-19 pandemic.
\newblock Journal of Public Policy \& Marketing. 2021;40(1):108-10.

\bibitem{burmester2007rfid}
Burmester M, De~Medeiros B.
\newblock RFID security: attacks, countermeasures and challenges.
\newblock In: The 5th RFID academic convocation, the RFID journal conference.
  Citeseer; 2007. .

\bibitem{ziavrou2022trends}
Ziavrou KS, Noguera S, Boumba VA.
\newblock Trends in counterfeit drugs and pharmaceuticals before and during
  COVID-19 pandemic.
\newblock Forensic Science International. 2022:111382.

\bibitem{ofori2022fighting}
Ofori-Parku SS.
\newblock Fighting the global counterfeit medicines challenge: A
  consumer-facing communication strategy in the US is an imperative.
\newblock Journal of Global Health. 2022;12.

\bibitem{counterfeit_pills_fact_sheet}
Counterfeit Pills Fact Sheet;.
\newblock Available from:
  \url{https://www.dea.gov/sites/default/files/2021-12/DEA-OPCK_FactSheet_December\%202021.pdf}.

\bibitem{krombholz2014qr}
Krombholz K, Fr{\"u}hwirt P, Kieseberg P, Kapsalis I, Huber M, Weippl E.
\newblock QR code security: A survey of attacks and challenges for usable
  security.
\newblock In: International Conference on Human Aspects of Information
  Security, Privacy, and Trust. Springer; 2014. p. 79-90.

\bibitem{tan2022review}
Tan WC, Sidhu MS.
\newblock Review of RFID and IoT integration in supply chain management.
\newblock Operations Research Perspectives. 2022:100229.

\bibitem{huang2020acd}
Huang W, Zhang Y, Feng Y.
\newblock ACD: An adaptable approach for RFID cloning attack detection.
\newblock Sensors. 2020;20(8):2378.

\bibitem{xu2018side}
Xu R, Zhu L, Wang A, Du X, Choo KKR, Zhang G, et~al.
\newblock Side-channel attack on a protected RFID card.
\newblock IEEE Access. 2018;6:58395-404.

\bibitem{kozicki2021secure}
Kozicki MN. Secure access with dendritic identifiers. Google Patents; 2021.
\newblock US Patent App. 17/100,028.

\bibitem{chi2020consistency}
Chi Z, Valehi A, Peng H, Kozicki M, Razi A.
\newblock Consistency penalized graph matching for image-based identification
  of dendritic patterns.
\newblock IEEE Access. 2020;8:118623-37.

\bibitem{li2020deep}
Li H, Chen X, Chi Z, Mann C, Razi A.
\newblock Deep DIH: single-shot digital in-line holography reconstruction by
  deep learning.
\newblock IEEE Access. 2020;8:202648-59.

\bibitem{chen2022dh}
Chen X, Wang H, Razi A, Kozicki M, Mann C.
\newblock DH-GAN: A Physics-driven Untrained Generative Adversarial Network for
  3D Microscopic Imaging using Digital Holography.
\newblock arXiv preprint arXiv:220512920. 2022.

\bibitem{kozicki2022dendritic}
Kozicki MN. Dendritic tags. Google Patents; 2022.
\newblock US Patent App. 17/311,154.

\bibitem{thiyaneswaran2020development}
Thiyaneswaran B, Anguraj K, Sindhu M, Yoganathan N, Jayanthi J.
\newblock Development of Iris Biological Features Extraction for Biometric
  Based Authentication to Prevent Covid Spread.
\newblock International Journal of Advanced Science and Technology.
  2020;29(3):8266-75.

\bibitem{kala2021contactless}
Kala H, Kumar S, Reddy RB, Shastry N, Thakur R.
\newblock Contactless Authentication Device using Palm Vein and Palm Print
  Fusion Biometric Technology for Post Covid World.
\newblock In: 2021 International Conference on Design Innovations for 3Cs
  Compute Communicate Control (ICDI3C). IEEE; 2021. p. 281-5.

\bibitem{ekiz2021end}
Ekiz D, Can YS, Darda{\u{g}}an YC, Aydar F, K{\"o}se RD, Ersoy C.
\newblock End-to-End Deep Multi-Modal Physiological Authentication With
  Smartbands.
\newblock IEEE Sensors Journal. 2021;21(13):14977-86.

\bibitem{kumari2021novel}
Kumari P, Seeja K.
\newblock A novel periocular biometrics solution for authentication during
  Covid-19 pandemic situation.
\newblock Journal of Ambient Intelligence and Humanized Computing.
  2021;12(11):10321-37.

\bibitem{wang2022fast}
Wang H, Chen X, Razi A.
\newblock Fast Key Points Detection and Matching for Tree-Structured Images.
\newblock arXiv preprint arXiv:221103242. 2022.

\bibitem{valehi2017graph}
Valehi A, Razi A, Cambou B, Yu W, Kozicki M.
\newblock A graph matching algorithm for user authentication in data networks
  using image-based physical unclonable functions.
\newblock In: 2017 Computing Conference. IEEE; 2017. p. 863-70.

\bibitem{lee2021cybersecurity}
Lee I.
\newblock Cybersecurity: Risk management framework and investment cost
  analysis.
\newblock Business Horizons. 2021;64(5):659-71.

\bibitem{kalender2006x}
Kalender WA.
\newblock X-ray computed tomography.
\newblock Physics in Medicine \& Biology. 2006;51(13):R29.

\bibitem{abrahamsson2015multifocus}
Abrahamsson S, McQuilken M, Mehta SB, Verma A, Larsch J, Ilic R, et~al.
\newblock MultiFocus Polarization Microscope (MF-PolScope) for 3D polarization
  imaging of up to 25 focal planes simultaneously.
\newblock Optics Express. 2015;23(6):7734-54.

\bibitem{rivenson2018phase}
Rivenson Y, Zhang Y, G{\"u}nayd{\i}n H, Teng D, Ozcan A.
\newblock Phase recovery and holographic image reconstruction using deep
  learning in neural networks.
\newblock Light: Science \& Applications. 2018;7(2):17141-1.

\bibitem{yang20173d}
Yang B, Wen H, Wang S, Clark R, Markham A, Trigoni N.
\newblock 3d object reconstruction from a single depth view with adversarial
  learning.
\newblock In: Proceedings of the IEEE international conference on computer
  vision workshops; 2017. p. 679-88.

\bibitem{amann20193d}
Amann S, Witzleben Mv, Breuer S.
\newblock 3D-printable portable open-source platform for low-cost lens-less
  holographic cellular imaging.
\newblock Scientific reports. 2019;9(1):1-10.

\end{thebibliography}
\bibliographystyle{vancouver}

\end{document}